\documentclass[3p,times,procedia]{elsarticle}
\usepackage{nupha_ecrc}

\volume{00}

\firstpage{1}

\journalname{Nuclear Physics A}

\runauth{}

\jid{nupha}

\jnltitlelogo{Nuclear Physics A}

\usepackage{amssymb}
\usepackage{amsmath}

\usepackage[figuresright]{rotating}

\begin{document}

\begin{frontmatter}



\dochead{XXVIth International Conference on Ultrarelativistic Nucleus-Nucleus Collisions\\ (Quark Matter 2017)}

\title{Quantifying the Chiral Magnetic Effect from Anomalous-Viscous Fluid Dynamics}


\author[IU]{Shuzhe Shi\footnote{Presenter}} 
\author[HB]{Yin Jiang}
\author[MC,IU]{Elias Lilleskov}
\author[MIT]{Yi Yin}
\author [IU]{Jinfeng Liao} 

\address[IU]{Physics Department and Center for Exploration of Energy and Matter,
Indiana University, 2401 N Milo B. Sampson Lane, Bloomington, IN 47408, USA.}
\address[HB]{Institut f\"ur Theoretische Physik, Universit\"at Heidelberg, Philosophenweg 16, D-69120 Heidelberg, Germany.}
\address[MC]{Department of Physics and Astronomy, Macalester College, 
1600 Grand Avenue, Saint Paul, MN 55105, USA.}
\address[MIT]{Center for Theoretical Physics, Massachusetts Institute of Technology, Cambridge, MA 02139, USA.}

\begin{abstract}
In this contribution we report a recently developed Anomalous-Viscous Fluid Dynamics (AVFD) framework, which simulates the evolution of fermion currents in QGP on top of  the bulk expansion from data-validated VISHNU hydrodynamics. With reasonable estimates of initial conditions and magnetic field lifetime, the predicted CME signal is quantitatively consistent with  change separation measurements in 200GeV Au-Au collisions at RHIC. We further develop the event-by-event AVFD simulations that allow 
direct evaluation of  two-particle correlations arising from  CME signal as well as the non-CME backgrounds. Finally we report predictions from AVFD simulations for the upcoming isobaric (Ru-Ru v.s. Zr-Zr ) collisions that could provide the critical test of the CME in heavy ion collisions.
\end{abstract}


\end{frontmatter}


\section{Introduction -- Chiral Magnetic Effect \& Anomalous-Viscous Fluid Dynamics}
\label{intro}
The Chiral Magnetic Effect (CME)~\cite{Kharzeev:2004ey,Kharzeev:2015znc} refers to the generation of an electric current    $\vec{\bf J}_Q$ along the {\it magnetic field} $\vec{\bf B}$ applied to a system of chiral fermions with chirality imbalance, i.e.
\vspace{-0.1in}
\begin{eqnarray} \label{eq_cme}
\vec{\bf J}_Q = \sigma_5 \vec{\bf B}
\end{eqnarray}
where $\sigma_5 = C_A \mu_5$ is the chiral magnetic conductivity, with the chiral chemical potential $\mu_5$ that quantifies the imbalance between fermion densities of opposite (right-handed, RH versus left-handed, LH) chirality. 

Given the magnificent physics embodied in the Chiral Magnetic Effect, it is of the utmost interest to search for its manifestation in the quark-gluon plasma (QGP) created in relativistic heavy ion collisions at the Relativistic Heavy Ion Collider (RHIC) and the Large Hadron Collider (LHC). Dedicated searches for potential CME signals have been ongoing at RHIC and the LHC~\cite{STAR_LPV1,STAR_LPV_BES,ALICE_LPV,Khachatryan:2016got}, with encouraging evidences reported through measuring the charge separation signal induced by the CME current (\ref{eq_cme}). The interpretation of these data however suffers from backgrounds arising from the complicated environment in a heavy ion collision (see e.g.~\cite{Kharzeev:2015znc,Liao:2014ava,Bzdak:2012ia}). Currently the most pressing challenge for the search of CME in heavy ion collisions is to clearly separate background contributions from the desired signal. A mandatory and critically needed step, is to develop state-of-the-art modeling tools that can quantify CME contribution in a realistic heavy ion collision environment. 

To address this challenge, we've recently developed a simulation framework, the Anomalous-Viscous Fluid Dynamics (AVFD) \cite{Jiang:2016wve}, focusing on describing anomalous chiral transport in heavy ion collisions at  high beam energy (such as the top energy RHIC collisions). The bulk evolution in such collisions is well described by boost-invariant 2+1D 2nd-order viscous hydrodynamics (e.g. VISHNU simulations~\cite{Shen:2014vra}) where net charge densities are small enough and typically neglected without much influence on bulk evolution. However to study the CME, one needs to accurately account for the evolution of fermion currents. Our approach is to solve the following fluid dynamical equations for the chiral fermion currents (RH and LH currents for u and d flavors respectively) as perturbations on top of the bulk fluid evolution: 
\vspace{-0.1in}
\begin{eqnarray} 
\begin{aligned}[c]
\hat{D}_\mu J_{\chi, f}^\mu &= \chi \frac{N_c Q_f^2}{4\pi^2} E_\mu B^\mu \\
J_{\chi, f}^\mu &= n_{\chi, f}\, u^\mu + \nu_{\chi, f}^\mu + \chi \frac{N_c Q_f}{4\pi^2} \mu_{\chi, f} B^\mu  \\ 
\end{aligned}\qquad\qquad
\begin{aligned}[c]
\Delta^{\mu}_{\,\, \nu} \hat{d} \left(\nu_{\chi, f}^\nu \right) &= - \frac{1}{\tau_{r}} \left[  \left( \nu_{\chi, f}^\mu \right) -  \left(\nu_{\chi, f}^\mu \right)_{NS} \right ] \\
\left(\nu_{\chi, f}^\mu \right)_{NS} &=  \frac{\sigma}{2} T \Delta^{\mu\nu}   \partial_\nu \left(\frac{\mu_{\chi, f}}{T}\right) +  \frac{\sigma}{2} Q_f E^\mu   \quad
\end{aligned}
 \label{eq_avfd}
\end{eqnarray} 
where $\chi=\pm1$ labels chirality for RH/LH currents and $f=u,d$ labels light quark flavor with  electric charge $Q_f$ and color factor $N_c=3$. The $E^\mu=F^{\mu\nu}u_\nu$ and  $B^\mu=\frac{1}{2}\epsilon^{\mu\nu\alpha\beta}u_\nu F_{\alpha\beta}$ are   external electromagnetic fields in fluid rest frame.  The derivative $\hat{D}_\mu$ is covariant derivative and  $\hat{d} =u^\mu \hat{D}_\mu$, with projection operator $\Delta^{\mu \nu}=\left(g^{\mu\nu} - u^\mu u^\nu \right)$.  Viscous parameters $\sigma$ and $\tau_r$ are diffusion constant and relaxation time respectively. 
In the above equations the fluid four-velocity field $u^\mu$, temperature $T$ as well as all other thermodynamic quantities are determined by background bulk flow. Furthermore the (small) fermion densities $n_{\chi, f}$ and corresponding chemical potential $ \mu_{\chi, f}$ are related by lattice-computed quark number susceptibilities $c_2^f(T)$.

\section{Results and Discussion}
\label{result}

As the key ingredients for driving the CME current, the magnetic field and the initial axial charge density are the most important inputs for the AVFD simulation. 
For the magnetic field $\vec {B} = B(\tau) \hat{y}$ (with $\hat{y}$ the event-wise out-of-plane direction), we use a plausible parametrization (see e.g.~\cite{Yin:2015fca}) as   
$B(\tau) = \frac{B_0}{1+\left(\tau / \tau_B\right)^2}$.
The peak value $B_0$ (for each centrality) at the collision point has been well quantified with event-by-event simulations and we use the most realistic values from \cite{Bloczynski:2012en}. For the lifetime of the $B$ field we use a reasonable estimate of $\tau_B=0.6$ fm/c which is  comparable to the onset time of hydrodynamic evolution. 
For the initial axial charge density arising from gluonic topological charge fluctuations, one could make the following estimate based on the strong chromo-electromagnetic fields in the early-stage glasma similarly to the recent study in \cite{Hirono:2014oda}: 
$ \sqrt{ \left< n_5^2 \right> } \simeq  \frac{Q_s^4\, (\pi \rho_{tube}^2 \tau_0) \, \sqrt{N_{coll.}}}{16\pi^2 \, A_{overlap}} $.
In the above $\rho_{tube}\simeq 1 \rm fm$ is the transverse extension of glasma flux tube, $A_{overlap}$ is the geometric overlapping area of the two colliding nuclei, and $N_{coll.}$ the binary collision number for a given centrality. Such axial charge density depends most sensitively upon the saturation scale $Q_s$, with a reasonable range of  $Q_s^2\simeq 1\sim 1.5 \rm GeV^2$ for RHIC 200AGeV collisions.

\begin{figure}[!hbt]
 \hspace{4pc}\includegraphics[width=0.36\textwidth]{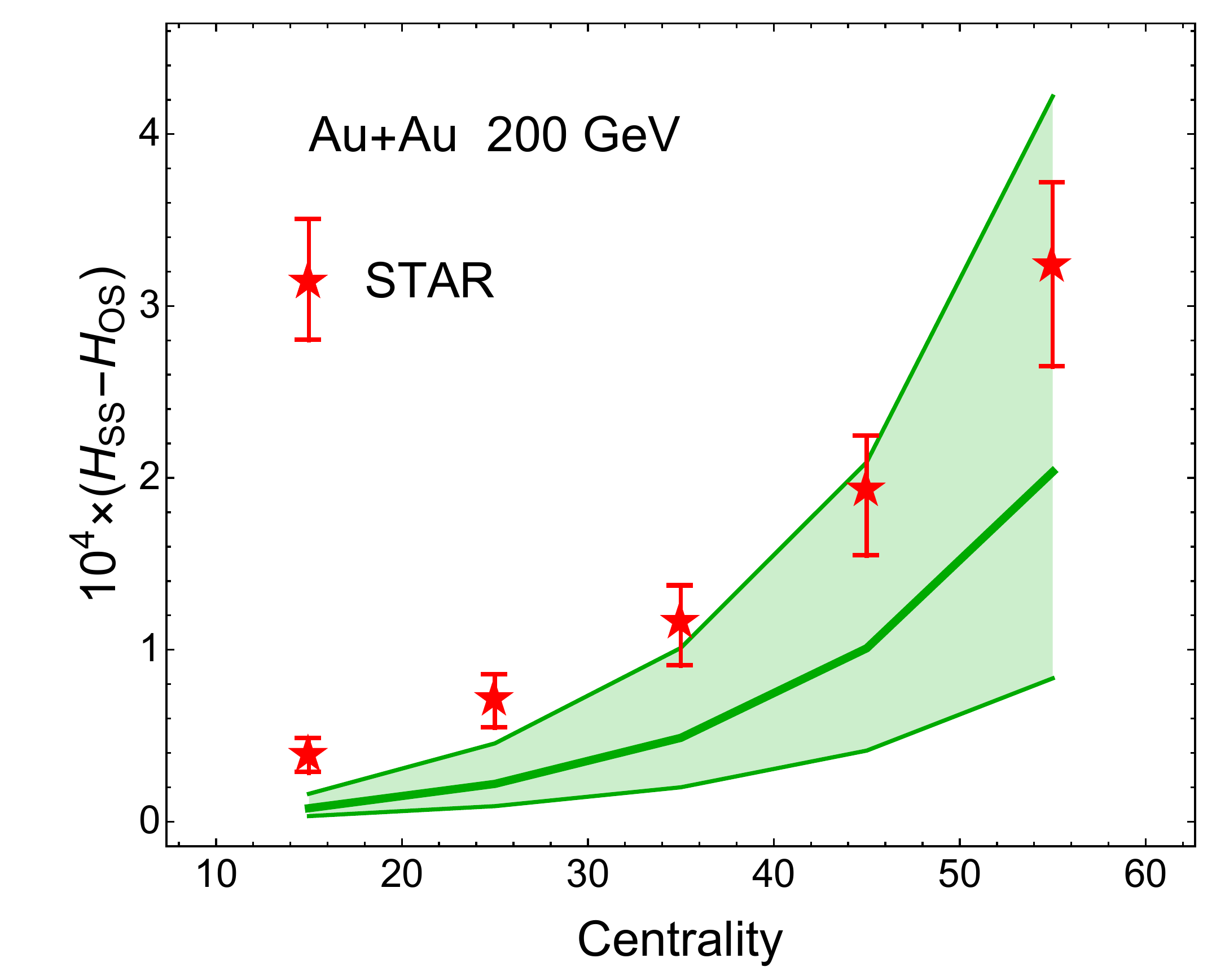} \hspace{2pc}%
\begin{minipage}[b]{10pc}
\begin{center} 
\caption{\label{fig_H} 
 Quantitative predictions from Anomalous-Viscous Fluid Dynamics simulations for the CME-induced H-correlations, in comparison with STAR measurements~\cite{STAR_LPV_BES}. The green bands reflect current theoretical uncertainty in the estimates of initial axial charge generated by gluonic field topological fluctuations.}
\end{center}
\end{minipage}\vspace{-0.1in}
\end{figure}

After the preceding discussions on the various aspects of the AVFD tool, let us now proceed to utilize this tool for quantifying CME signal to be compared with available data. The measurement of a CME-induced charge separation is however tricky, as this dipole flips its sign  from event to event depending the sign of the initial axial charge arising from fluctuations, thus with a vanishing event-averaged mean value. What can be measured is its variance, through azimuthal correlations for same-sign (SS) and opposite-sign (OS) pairs of charged hadrons. The so-called $\gamma_{SS/OS}\equiv \left< \cos(\phi_1+\phi_2)\right>$ observables measure a difference between the in-plane versus out-of-plane correlations and are indeed sensitive to potential CME contributions. They however suffer from considerable flow-driven background contributions that are not related to CME (see e.g. \cite{Kharzeev:2015znc,Bzdak:2012ia}). One plausible approach to  separate background and CME signal is based on a two-component scenario~\cite{Bzdak:2012ia}, which  was recently adopted by the STAR Collaboration to suppress backgrounds and   extract the flow-independent part (referred to as $H_{SS/OS}$)~\cite{STAR_LPV_BES}. We consider $H_{SS/OS}$ as our ``best guess'' thus far for potential CME signal to be compared with AVFD computations. Specifically a pure CME-induced charge separation will contribute as $\left(H_{SS}-H_{OS}\right) \to 2\left(a^{ch}_1\right)^2$. The AVFD results for various centrality bins are presented in Fig.~\ref{fig_H}, with the green band spanning the range of key parameter $Q_s^2$ in $1\sim 1.5 \rm GeV^2$  reflecting uncertainty in estimating the initial axial charge. Clearly the CME-induced correlation is very sensitive to the amount of initial axial charge density as controlled by $Q_s^2$. The comparison with STAR data~\cite{STAR_LPV_BES} shows very good agreement for the magnitude and centrality trend for choices with relatively large values of $Q_s^2$. 

\begin{figure*}[!hbt]
\begin{center} \vspace{-0.15in}
\includegraphics[width=0.32\textwidth]{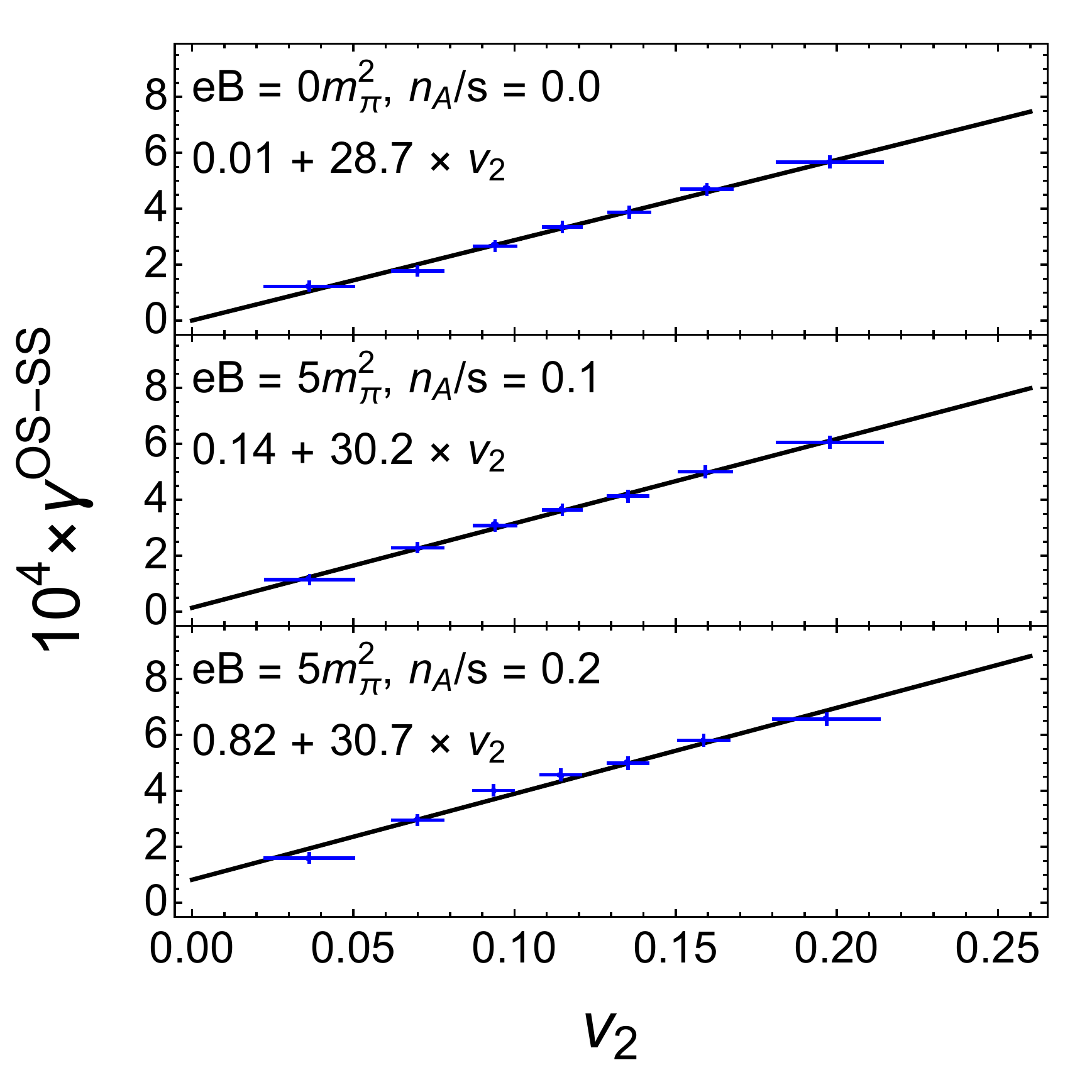} \hspace{0.4in}
\includegraphics[width=0.32\textwidth]{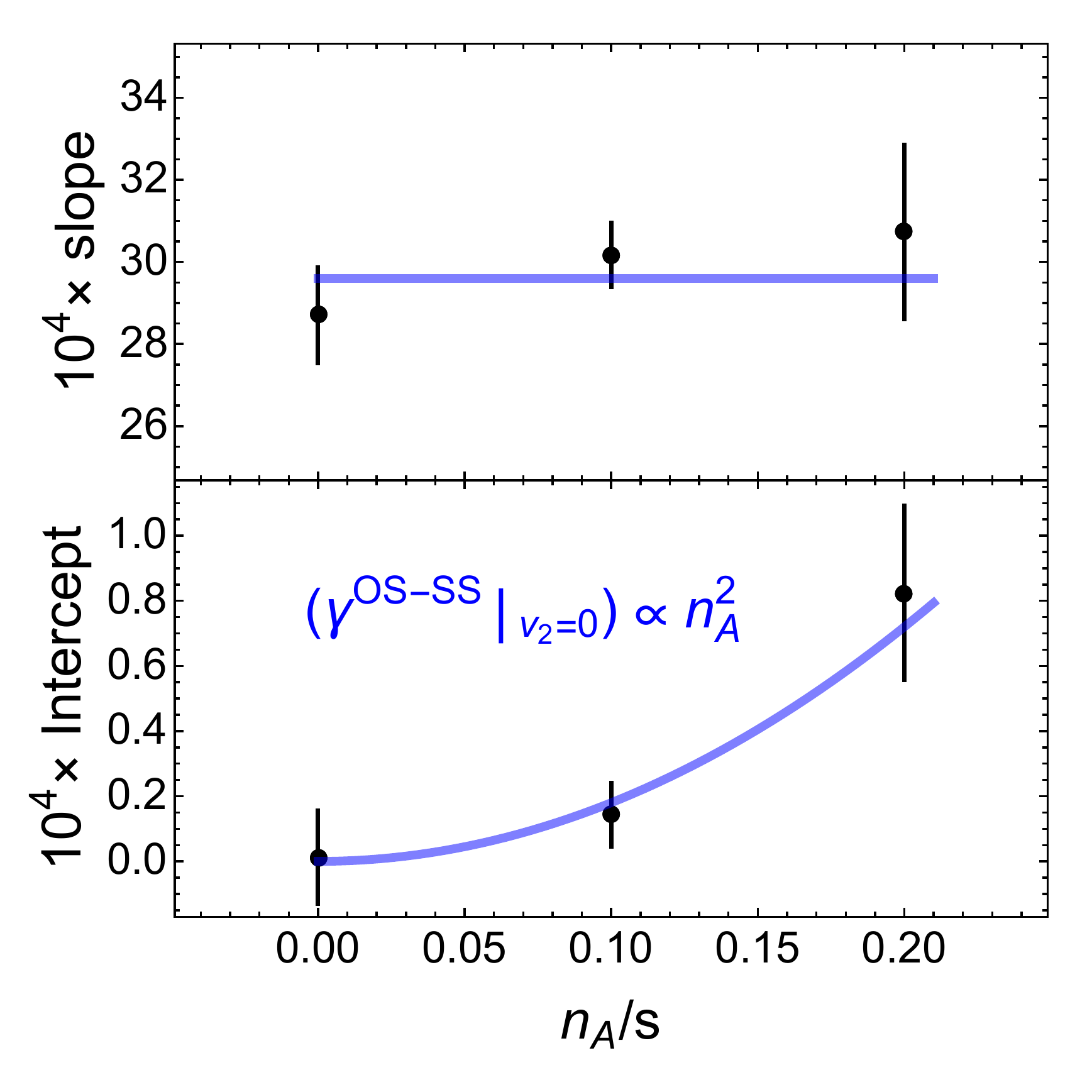}
\vspace{-0.15in}
\caption{(color online) (left) The $v_2$ dependence of $\gamma^{OS-SS}$; (right)  The slope and intercept extracted from the $v_2$ dependence of $\gamma^{OS-SS}$. }  \label{fig_gamma}
\end{center}
\vspace{-0.2in}
\end{figure*}

While the AVFD simulations based on smooth average hydro profile provide a quantitative account of the ``pure" CME signal, it is imperative to perform event-by-event simulations that could allow direction evolution of two-particle correlations and thus ultimate comparison with experimental measurements. We've recently implemented such simulations with event-wise fluctuating initial conditions as well as a hadronic stage after hadronization using UrQMD  to include the effects of hadron cascade and resonance decay. To demonstrate the CME-driven correlations, we  run simulations for $50-60\%$ 200GeV Au-Au collisions with three different setups: (a) a ``null'' case with no magnetic field or chiral imbalance; (b) $eB=5m_\pi^2$ and $n_A/s=0.1$ (corresponding to estimates with $Q_s^2\simeq 1 \rm GeV^2$); (c) $eB=5m_\pi^2$ and $n_A/s=0.2$ (corresponding to estimates with $Q_s^2\simeq 1.5 \rm GeV^2$). For each setup, we accumulated $\sim10^{7}$ events and measured the correlator $\gamma^{OS-SS}$ versus different event-wise $v_2$ bins, shown in Fig.\ref{fig_gamma} (left). An obvious linear dependence $\gamma^{OS-SS}$ on $v_2$ is observed for all these three cases. The extracted slope and intercept of such linear dependence are shown in Fig.\ref{fig_gamma} (right): while  the slope is basically independent of initial axial charge (indicating its dominant non-CME origin), the intercept grows quadratically with initial axial charge as expected from the CME signal.

Finally we report the AVFD predictions for the isobaric collisions planned at RHIC~\cite{Skokov:2016yrj}, which provide a unique means to unambiguously decipher CME from backgrounds. In such ``contrast'' colliding systems (specifically for $^{96}_{40}$Zr-$^{96}_{40}$Zr versus $^{96}_{44}$Ru-$^{96}_{44}$Ru), the bulk particle production and background correlations are expected to be basically identical while the CME signals are expected to differ by about $20\%$ due to their different magnetic field strength by about $10\%$. This will be a crucial test for the search of CME, and quantitative predictions are important. 
\begin{figure*}[!hbt]
\begin{center} \vspace{-0.1in}
\includegraphics[width=0.32\textwidth]{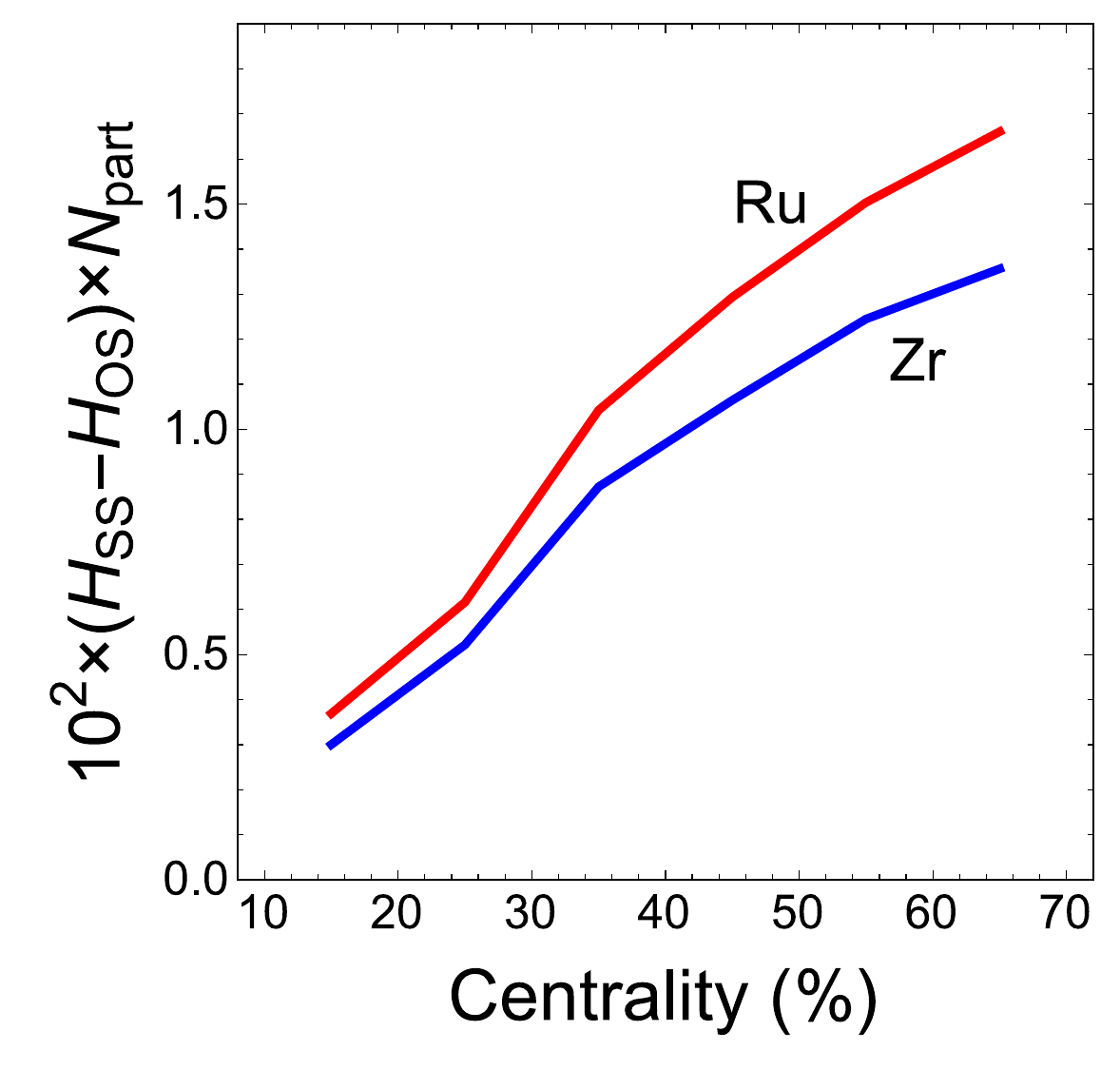} \hspace{0.02in}
\includegraphics[width=0.32\textwidth]{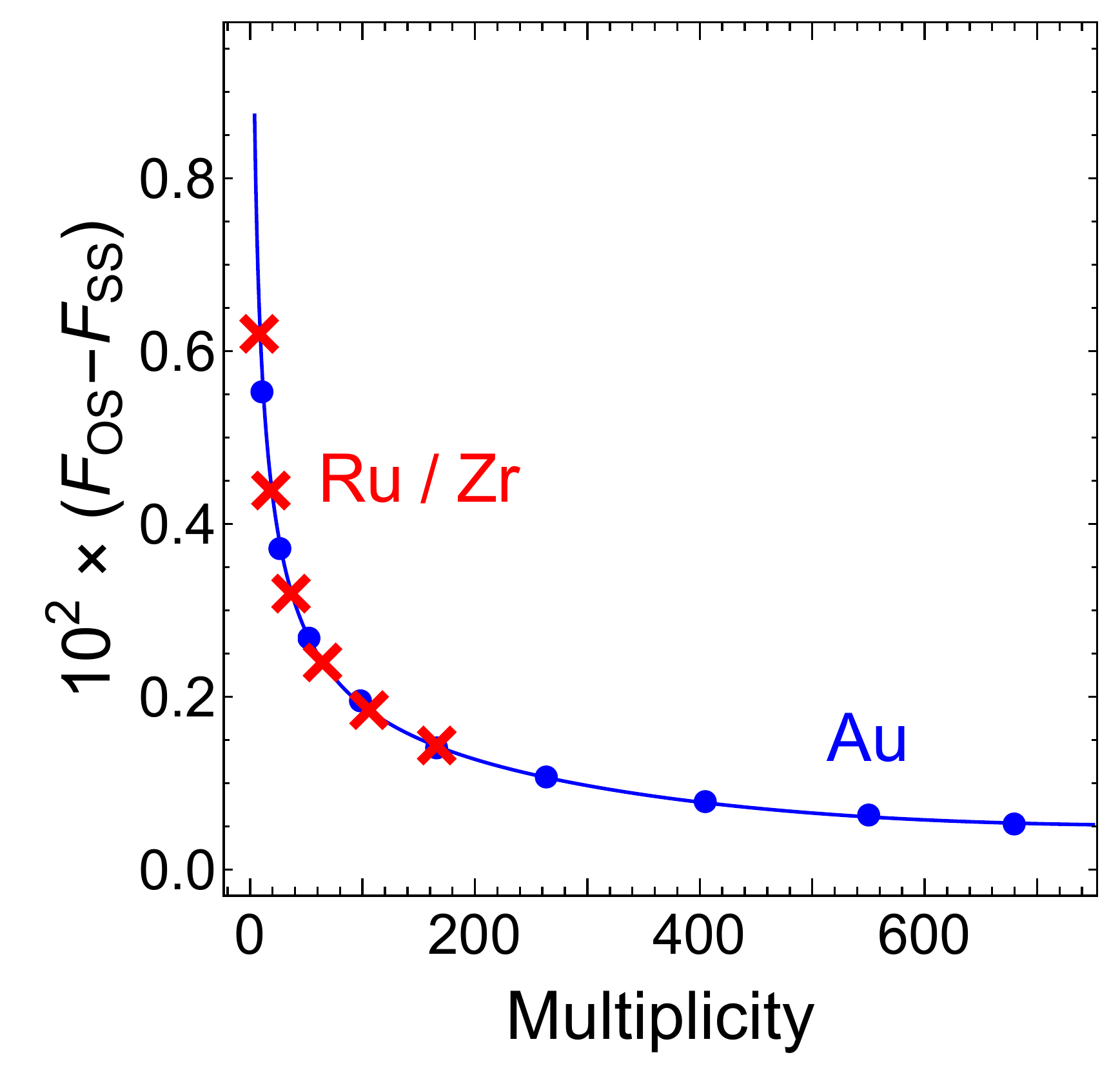} \hspace{0.02in}
\includegraphics[width=0.32\textwidth]{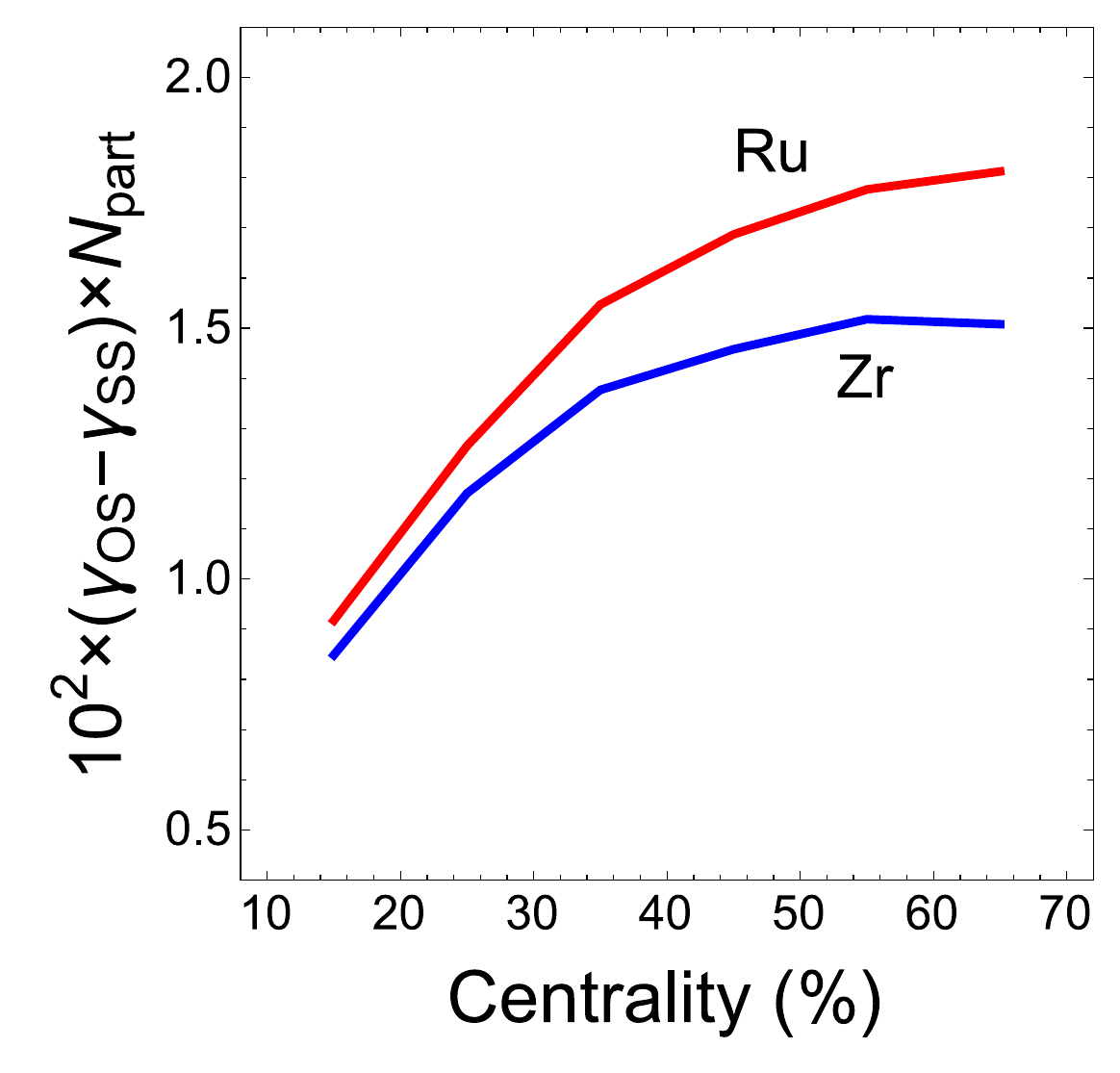}
\vspace{-0.15in}
\caption{(color online) (left) AVFD predictions for CME-induced H-correlations in isobar collisions; (middle) Extrapolation of background correlations from AuAu to ZrZr and RuRu systems; (right) Predicted $\gamma$-correlations in ZrZr and RuRu collisions by folding together computed CME signals and estimated background correlations. }  \label{fig_iso}
\end{center}
\vspace{-0.2in}
\end{figure*} 
In Fig.~\ref{fig_iso} we show the AVFD results for the correlations in isobar collisions. The left panel shows our prediction for pure CME signal in these two systems. By assuming the background correlation strength is mainly dependent on multiplicity, we extrapolate the departed background strength from Au-Au to the isobaric collisions (shown in middle panel). Folding CME signal with such background estimates we obtain the predictions for $\gamma$-correlations, with a visible $\sim 15\%$ shift in peripheral region differentiating the two systems. Provided current experimental uncertainty and limitations\cite{Deng:2016knn}, the difference at such a level should be measurable with adequate number of events from isobaric collisions.

\section{Summary}
\label{sum}
In summary, a new simulation tool --- the Anomalous-Viscous Fluid Dynamics (AVFD) framework has been developed for quantifying the charge separation signal induced by Chiral Magnetic Effect in relativistic heavy ion collisions. We find that,   
subject to current theoretical and experimental uncertainties, the AVFD-predicted CME signal with realistic initial conditions and magnetic field lifetime is quantitatively consistent with measurements from 200AGeV AuAu collisions at RHIC. Also, by event-by-event simulation, we find that the intercept of the $v_2$ dependence of the correlator $\gamma^{SS-OS}$ is directly   sensitive to the CME while the slope is likely to be dominated by background contributions.
Finally, we make predictions for the upcoming isobaric collisions that would be a critical test for the search of CME in heavy ion collisions.

 {\bf Acknowledgments.} 
This material is based upon work supported by the U.S. Department of Energy, Office of Science, Office of Nuclear Physics, within the framework of the Beam Energy Scan Theory (BEST) Topical Collaboration (YJ, JL, SS and YY). The work is also supported in part by the NSF Grant No. PHY-1352368 (JL and SS), by the U.S. DOE grant Contract Number No. DE-SC0011090 (YY), by the DFG Collaborative Research Center  ``SFB 1225 (ISOQUANT)'' (YJ), and by the NSF REU Program at Indiana University (EL).





\bibliographystyle{elsarticle-num}



\end{document}